\documentclass[preprint]{imsart}
\RequirePackage{amsthm,amsmath,amsfonts,amssymb}
\RequirePackage[authoryear]{natbib}
\RequirePackage[colorlinks,citecolor=blue,urlcolor=blue]{hyperref}
\RequirePackage{graphicx, commath}
\usepackage{bm}
 
\startlocaldefs
\theoremstyle{plain}

\theoremstyle{definition}

\endlocaldefs

\begin{document}

\begin{frontmatter}
\title{Joint Group-Based Trajectory Modeling for Paired Repeated Measures: An Application to Audiometric Phenotypes and Dietary Associations}
\runtitle{Joint GBTM for Paired Repeated Measures}

\begin{aug}
%%%%%%%%%%%%%%%%%%%%%%%%%%%%%%%%%%%%%%%%%%%%%%%
%% Only one address is permitted per author. %%
%% Only division, organization and e-mail is %%
%% included in the address.                  %%
%% Additional information such as            %%
%% identifying the corresponding author must %%
%% be included in in the Acknowledgments     %%
%% section if necessary.                     %%
%% ORCID can be inserted by command:         %%
%% \orcid{0000-0000-0000-0000}               %%
%%%%%%%%%%%%%%%%%%%%%%%%%%%%%%%%%%%%%%%%%%%%%%%
\author[A]{\fnms{Ying}~\snm{Chen} \ead[label=e1]{ying\_chen@hsph.harvard.edu}}
\author[B, C]{\fnms{Sharon}~\snm{Curhan}\ead[label=e2]{scurhan@bwh.harvard.edu}}
\author[D]{\fnms{Kenneth}~\snm{I. Vaden Jr}\ead[label=e3]{vaden@musc.edu}}
\author[D]{\fnms{Judy}~\snm{R. Dubno}\ead[label=e4]{dubnojr@musc.edu}}
\author[A,B,C,E]{\fnms{Molin}~\snm{Wang}\ead[label=e5]{stmow@channing.harvard.edu}}
%%%%%%%%%%%%%%%%%%%%%%%%%%%%%%%%%%%%%%%%%%%%%%
%% Addresses                                %%
%%%%%%%%%%%%%%%%%%%%%%%%%%%%%%%%%%%%%%%%%%%%%%
\address[A]{Department of Epidemiology, Harvard T.H. School of Public Health \printead[presep={ ,\ }]{e1}}

\address[B]{Harvard Medical School \printead[presep={,\ }]{e2}}

\address[C]{Department of Medicine, Brigham and Womens Hospital \printead[presep={,\ }]{}}

\address[D]{Department of Otolaryngology-Head and Neck Surgery, Medical University of South Carolina\printead[presep={,\ }]{e3,e4}}

\address[E]{Department of Biostatistics, Harvard T.H. School of Public Health \printead[presep={,\ }]{e5}}
\end{aug}

\begin{abstract}
%Conventional group-based trajectory modeling (GBTM) provides a flexible framework for identifying distinct patterns in repeated-measures data, but generally assumes conditional independence given latent group membership. 
The assumption of conditional independence  in conventional group-based trajectory modeling (GBTM) is often violated by paired repeated-measures data with heterogeneous trajectory patterns.
While random-effects models can accommodate this dependence, they inflate within-group variability and blur distinct phenotypic shapes. We propose a joint GBTM framework that explicitly models hierarchical dependence in paired trajectories while allowing them to follow different latent patterns. We develop a robust two-stage approach to address estimation challenges caused by rare latent groups, and a one-stage EM algorithm that serves as a theoretical baseline under balanced group sizes. Simulations demonstrate that our methods correct the biases caused by ignoring hierarchical dependence.  The proposed model was applied to real-world data from the Conservation of Hearing Study (CHEARS) Audiology
Assessment Arm (AAA), a subcohort of the Nurses’ Health Study II (NHS II),  to identify distinct audiometric phenotypes and to investigate the association between the Dietary Approaches to Stop Hypertension (DASH) dietary adherence score and the latent audiometric patterns.
\end{abstract}

\begin{keyword}
\kwd{Group-based trajectory modeling}
\kwd{paired repeated measurements}
\kwd{mixture modeling}
\kwd{multilevel correlation}
\kwd{audiometric multilevel clustering}
\end{keyword}

\end{frontmatter}
%%%%%%%%%%%%%%%%%%%%%%%%%%%%%%%%%%%%%%%%%%%%%%
%% Please use \tableofcontents for articles %%
%% with 50 pages and more                   %%
%%%%%%%%%%%%%%%%%%%%%%%%%%%%%%%%%%%%%%%%%%%%%%
%\tableofcontents

\section{Introduction}

Identifying distinct patterns in repeated measurement data is essential for understanding disease progression and population heterogeneity in biological and medical research. The latent class growth analysis or group-based trajectory modeling (GBTM)  \citep{nagin1999analyzing, jones2001sas,nagin2005group}, a widely used finite mixture model-based framework in clinical research, provides a flexible framework to capture this heterogeneity by uncovering latent groups with distinct trajectory patterns. Comprehensive reviews of GBTM are available in \cite{nagin2010group2,nagin2010group1}. 
 Despite their flexibility,
conventional GBTMs generally rely on the assumption of conditional independence  of trajectories given latent group membership. This assumption may not hold in the context of paired repeated measurement data, such as when a single participant contributes two potentially correlated trajectories from paired organs (e.g., the two ears, kidneys, or lungs). Ignoring this inherent hierarchical covariance and applying GBTM  naively can result in biased parameter estimates, trajectory misclassification, and misleading inferences regarding the associations between exposures and latent group memberships \citep{goldstein2011multilevel, davies2017impact}. 

Our work is motivated by the Conservation of Hearing Study (CHEARS), which investigates risk factors
for acquired hearing loss  among participants in several large ongoing cohorts, including the Nurses’ Health Study II (NHS II). In 1989, NHS II enrolled $116,430$ US female registered nurses  aged 25--42 years in 1989 \citep{curhan2018adherence}. In a subcohort of the NHS II, the CHEARS Audiology Assessment Arm (AAA), 3136 participants have undergone clinical audiometric assessments in 2012--2014 \citep{curhan2020prospective}, with air-conduction hearing thresholds measured across seven pure-tone frequencies ($500,1000,2000,3000,4000,6000,8000$ Hz) in both ears. The CHEARS-AAA hearing threshold data exhibit a two-level hierarchical dependence structure: hearing thresholds across different frequencies are correlated within each ear, and the two ears are correlated within the same participant through both same-frequency and cross-frequency dependence due to shared participant-level risk factors.
In this study, we aim to identify distinct audiometric patterns based on hearing thresholds across pure-tone  frequencies in both ears, and to assess their  associations with the Dietary Approaches to
Stop Hypertension (DASH) dietary adherence score. In related applications, paired data arising from such multilevel clustering are often analyzed at one level due to the lack of appropriate methods. For example, previous studies have applied supervised learning \citep{dubno2013classifying, vaden2017longitudinal} or Gaussian mixture models \citep{parthasarathy2020data} to classify audiometric patterns for the left and right ears separately. While recent approaches attempt to address between-ear dependence \citep{yang2024soft}, they rely on predefined phenotypes and only account for the correlation marginally after classification.
By overlooking the full joint hierarchical correlation structure inherent in paired repeated measurements, these existing methods may lead to information loss and potentially biased inference.

Existing trajectory analyses handling multilevel dependence frequently utilize Growth Mixture Models (GMM) \citep{muthen1999finite, proust2017estimation}, the inclusion of multilevel random effects, however, assumes the existence of literal subpopulations with natural within-group variation \citep{ng2014mixture, sheng2022analytical}. When the objective is to approximate a continuous distribution of trajectories into strict prototypical groups, GBTM is conceptually more appropriate than GMM \citep{nagin2010group1}. 
 For the CHEARS-AAA data, GBTM aligns directly with our clinical objective because it strictly separates these audiometric trajectories into distinct phenotypic groups, without allowing  random effects to obscure the underlying patterns. 
Nevertheless, existing GBTM extensions are ill-equipped for paired trajectory data. Group-based dual-trajectory modeling \citep{nagin2001analyzing} accommodates related trajectories, but it models dependence merely via linking probabilities between independent group assignments. %, failing to leverage the hierarchical dependent structure of paired trajectories. 
Conversely, group-based multi-trajectory modeling \citep{nagin2018group} establishes a joint profile, but it forces both paired trajectories to share a single latent group assignment and relies on the assumption of conditional independence between them. To our knowledge, no GBTM framework currently exists that can simultaneously (i) handle the hierarchical correlation inherent in paired trajectories, and (ii) jointly model the paired trajectories while permitting them to be assigned to different latent groups.

To address the aforementioned methodological gap and  appropriately account for the hierarchical dependence in the CHEARS-AAA data, we propose a joint GBTM framework %to address the aforementioned methodological gap by incorporating
that incorporates residual covariance matrix while allowing paired trajectories to be assigned to different latent groups. While GBTM is typically applied to longitudinal data to model changes over time, the method is directly applicable to data measured across any ordered domain \citep{nagin2005group}. In our analysis of the CHEARS-AAA data, we define trajectories over an ordered sequence of frequencies rather than time, constructing audiogram trajectories to characterize and classify clinical patterns of hearing loss. Although the joint modeling approach is conceptually appealing, its specific formulation presents both methodologically and computationally challenges. Given that paired trajectories can belong to different phenotypes, relaxing the assumption of conditional independence requires maximizing a complex joint likelihood function across all possible combinatorial group assignments, which poses optimization challenges and identifiability issue, particularly when rare latent groups are present. To make this framework practically feasible,
  we provide two likelihood-based estimation procedures. First, we propose a computationally robust two-stage approach to address identifiability challenges caused by rare latent groups. Specifically, we fit a covariate-free mixture to obtain group membership assignments probabilities, then link these probabilities to exposures via a mixed-effects multinomial logistic model. Second, we develop a statistically efficient one-stage approach as a theoretical baseline that achieves full maximum likelihood estimation, recommended for studies with balanced group sizes and adequate sample sizes.  %Each procedure offers distinct advantages in terms of statistical and computational efficiency. 
  
  The rest of this paper is organized as follows: In Section 2, we introduce the notation of GBTM, outline the proposed model, and in Section 3, we describe two iterative estimation approaches and implementation details. In Section 4, the proposed approaches are validated through simulation studies and in Section 5, these approaches are applied to the
CHEARS-AAA data. Section 6 concludes with a brief discussion.

\section{Methods}
\subsection{GBTM for paired repeated measurements}
We present the proposed GBTM for paired repeated measurements using the CHEARS-AAA dataset as an illustration.
Suppose we observe paired hearing threshold measurements $\bm{Y}_i=(\bm{Y}_{i1}^\top,  \bm{Y}_{i2}^\top)^\top$, where $\bm{Y}_{ij}=(Y_{ij1}, \dots, Y_{ijQ})^\top$ denotes a  vector of thresholds across frequencies with $Y_{ijq}$  the hearing measurement for ear $j$ of participant $i$ at frequency $h_q$ for $q=1,\dots,Q$, $j=1,2$, $i=1,\dots,n$. Let $K_{ij} \in \{1,\dots,K\}$ denote the latent group membership for ear $j$ of participant $i$, and $\bm{\varphi}_k=(\varphi_{k0},\dots,\varphi_{kp})^\top$ is a vector of basis function coefficients  for group $k$.
Conditional on the joint group memberships $(K_{i1},K_{i2})=(k_1, k_2)$, where $k_1$ and $k_2$ are not necessarily identical, we assume the joint model for the paired trajectories $\bm{Y}_{i}$:
\begin{equation*}
    \bm{Y}_{i}|_{(k_1, k_2)}=(\bm{I}_2 \otimes \bm{W})\bm{\varphi}_{k_1,k_2} + \bm{\varepsilon}_{i}, 
\end{equation*}
where $\bm{\varphi}_{k_1,k_2}=(\bm{\varphi}_{k_1}^\top,\bm{\varphi}_{k_2}^\top)^\top$,  $\bm{\varepsilon}_{i} = (\bm{\varepsilon}_{i1}^\top, \bm{\varepsilon}_{i2}^\top)^\top$ is the error terms, $\bm{I}_2$ is the 2-dimensional identity matrix, and $\bm{W}$ is a design matrix constructed from functional basis terms of the ordered frequencies thresholds $h_1,\dots,h_Q$, which is specified as
\begin{equation*}
    \bm{W}= \begin{pmatrix}
        1 & h_1 & g^\top(h_1) \\
        1 & h_2 & g^\top(h_2) \\
        \vdots & \vdots &\vdots \\
        1 & h_Q & g^\top(h_Q)
    \end{pmatrix},
\end{equation*}
with $g(h_q)$ representing a vector of nonlinear terms of $h_q$. 
We assume the error vector $\bm{\varepsilon}_{i} \sim \mathrm{MVN}(\bm{0},\bm{\Sigma}_{k_1k_2})$ conditional on group assignments $(k_1,k_2)$, and $\bm{\Sigma}_{{k_1k_2}}^\top=\bm{\Sigma}_{{k_1k_2}}$. We further decompose 
\begin{equation}
    \bm{\Sigma}_{k_1k_2} = (\bm{I}_2 \otimes \bm{V})\bm{\Lambda}_{k_1k_2}(\bm{I}_2 \otimes \bm{V}),
\label{sigma} 
\end{equation}
 where $\bm{V}$ is a $Q \times Q$ diagonal matrix of standard deviations across the $Q$ frequencies, and  $\bm{\Lambda}_{k_1k_2}$ is the  $2Q \times 2Q$ joint correlation matrix for the paired trajectories.

\subsection{Model for the joint mixing probabilities}
We let $\pi_{k}$ denote the marginal assignment probability of group $k$, i.e., $P(K_{ij}=k)=\pi_{k}$, and assume a covariate-dependent model for $\pi_{k}$ as in the mixture-of-experts model \citep{jordan1994hierarchical}.  Let $\bm{X}_{ij}=[1,\bm{x}_{ij}^\top]^\top$, where $\bm{x}_{ij}$ is a vector of risk factors associated with ear $j$ of participant $i$.  The  mixing proportion of a single ear $j$ is,
\begin{equation}
    \pi_{k}(\bm{X}_{ij}|\bm{b}_i) = \frac{\exp\{\bm{X}_{ij}^\top\bm{\beta}_{k}+b_{ik}\}}{1+\sum_{\ell=2}^{K} \exp\{\bm{X}_{ij}^\top\bm{\beta}_{\ell}+b_{i\ell}\}},  \mbox{ for } k = 2,\dots, K,
    \label{logit} \notag
\end{equation}
and $\pi_{1}(\bm{X}_{ij}|\bm{b}_i)=1-\sum_{\ell=2}^{K}\pi_{\ell}(\bm{X}_{ij}|\bm{b}_i)$, where $\bm{\beta}_k$ is a vector of unknown logistic regression coefficients for group $k$,
$\bm{b}_i=(b_{i2},\dots,b_{iK})^\top\sim \mathrm{MVN}(\bm{0},\bm{\Gamma})$ is an unobservable vector of random effects with  covariance matrix $\bm{\Gamma}=\mathrm{diag}\{\gamma_2,\dots,\gamma_K\}$. Conditional on $\bm{b}_i$, we define the joint mixing proportion $\pi_{k_1k_2}(\bm{X}_{ij}|\bm{b}_i)=P(K_{i1}=k_1,K_{i2}=k_2|\bm{b}_i)$ for phenotype pair $(k_1,k_2)$ is specified as $$\pi_{k_1k_2}(\bm{X}_i|\bm{b}_i)=  \pi_{k_1}(\bm{X}_{i1}|\bm{b}_i)\pi_{k_2}(\bm{X}_{i2}|\bm{b}_i).$$
%\cyan{While the group assignments for the two trajectories are conditionally independent given $\bm{b}_i$, shared participant-level random effects explicitly account for the inherent correlation in group memberships of paired trajectories.}

\subsection{The complete-data likelihood function}
We assume that the paired trajectories $\bm{Y}_i$ follow from a mixture model with $K \times K$ group pairs,
\begin{equation*}
    \bm{Y}_i \sim \sum_{k_1=1}^K \sum_{k_2=1}^K  \pi_{k_1 k_2}f_{k_1k_2}(\bm{Y}_{i};(\bm{I}_2 \otimes \bm{W})\bm{\varphi}_{k_1,k_2},\bm{\Sigma}_{k_1k_2}),
\end{equation*}
where  $f_{k_1k_2}(\bm{Y}_{i};(\bm{I}_2 \otimes \bm{W})\bm{\varphi}_{k_1,k_2},\bm{\Sigma}_{k_1k_2})$ is the multivariate normal density function with the mean vector $(\bm{I}_2 \otimes \bm{W})\bm{\varphi}_{k_1,k_2}$ and variance matrix $\bm{\Sigma}_{k_1k_2}$ conditional on joint group $(k_1,k_2)$.

Let $\bm{\Theta}=\{\bm{\beta},\bm{\varphi},\bm{\Gamma}, \bm{\Sigma}\}$ denote the full set of unknown parameters, where $\bm{\beta}=(\bm{\beta}_{2}^\top, \dots, \bm{\beta}_{K}^\top)^\top$, $\bm{\varphi}=(\bm{\varphi}_{1}^\top, \dots, \bm{\varphi}_{K}^\top)^\top$, and $\bm{\Sigma}= \{\bm{\Sigma}_{k},\bm{\Sigma}_{k_1k_2};k,k_1,k_2=1,\dots,K\}$. Given $n$ independent observations $\bm{\mathcal{Y}} = \{(\bm{X}_i,\bm{Y}_i); i=1,\dots,n\}$, we propose the following log-likelihood function for all $n$ paired trajectories:
\begin{equation}
    {L}(\bm{\Theta};\bm{\mathcal{Y}})=\sum_{i=1}^n
     \log \sum_{k_1=1}^K \sum_{k_2=1}^K \int_{\bm{b}_i}\pi_{k_1 k_2}(\bm{X}_i|\bm{b}_i)f_{k_1k_2}(\bm{Y}_{i};(\bm{I}_2 \otimes \bm{W})\bm{\varphi}_{k_1,k_2},\bm{\Sigma}_{k_1k_2}) f(\bm{b}_i;\bm{0},\bm{\Gamma})\dif \bm{b}_i,
    \label{likelihood}
\end{equation} where  $f(\bm{b}_i)$ is the normal density function of random effects $\bm{b}_i$.

  We introduce latent variables $Z_{ik_1k_2}=I(K_{i1}=k_1,K_{i2}=k_2)$ and let $\bm{Z}=\{Z_{ik_1k_2};k_1,k_2=1,\dots,K, i=1,\dots,n\}$ denote the set of all latent variables. By treating phenotype pair $(K_{i1},K_{i2})$ as incomplete data, the complete log-likelihood function for all $n$ paired trajectories is given by 
\begin{align}
    {L}^c(\bm{\Theta};\bm{\mathcal{Y}},\bm{Z})=& \sum_{i=1}^n \sum_{k_1=1}^K \sum_{k_2=1}^K  Z_{ik_1k_2} \log f_{k_1k_2}(\bm{Y}_{i};(\bm{I}_2 \otimes \bm{W})\bm{\varphi}_{k_1,k_2},\bm{\Sigma}_{k_1k_2}) \notag   \\
  &~~~~~~~~~~~~~~+ \sum_{i=1}^n \left[ \sum_{k_1=1}^K \sum_{k_2=1}^K  Z_{ik_1k_2} \log \int_{\bm{b}_i}\pi_{k_1 k_2}(\bm{X}_i|\bm{b}_i;\bm{\beta}) f(\bm{b}_i;\bm{0},\bm{\Gamma})\dif \bm{b}_i\right]. 
  \label{comp-likelihood}
\end{align}

\subsection{Hierarchical dependence structure in CHEARS-AAA data}
In the CHEARS-AAA data, hearing thresholds are correlated both within and between ears. Within each ear, thresholds at neighboring frequencies tend to move together; between ears, correlations arise because both ears share common risk factors (e.g., noise exposure,  systemic disease). To accommodate this hierarchical correlation, we consider the following exchangeable structures for both within-ear and between ear correlations. We assume $\rho_1^{(k)}$ is the correlation coefficient between two frequencies within an ear assigned to phenotype $k$, $\rho_2^{(k_1,k_2)}$ is between-ear correlation coefficient at the same frequency for ears assigned to phenotypes $k_1$ and $k_2$, and the correlation coefficient for different frequencies across the two ears is given by $\rho_2^{(k_1, k_2)}\rho_3^{(k_1, k_2)}$, with $\rho_3^{(k_1, k_2)}$ reflecting the reduction in correlation for different-frequency pairs between two ears. All correlation coefficients satisfy $0 < \rho_1^{(k)},\ \rho_2^{(k_1,k_2)},\ \rho_3^{(k_1,k_2)} < 1\ $, for $k, k_1, k_2 = 1, \ldots, K$. Let $\bm{\Lambda}_{k}$ be the within-ear correlation matrix when a single ear is assigned to phenotype $k$, and  $\bm{C}_{k_1k_2}$ the between-ear correlation matrix when the two ears are assigned to phenotypes $k_1$ and $k_2$. We have $\bm{\Lambda}_{k_1k_2}=\begin{pmatrix}
    \bm{\Lambda}_{k_1} & \bm{C}_{k_1k_2}\\\bm{C}_{k_1k_2} &\bm{\Lambda}_{k_2}
\end{pmatrix}$, and
\begin{equation}     \bm{\Lambda}_{k}=\begin{pmatrix}
        1 & \rho_1^{(k)} &\dots& \rho_1^{(k)}\\
        \rho_1^{(k)} & 1 & \dots & \rho_1^{(k)}\\
        \vdots & & \ddots & \vdots\\
        \rho_1^{(k)} &\rho_1^{(k)} &\dots&1
    \end{pmatrix}, \mbox{ } \bm{C}_{k_1k_2}=\rho_2^{(k_1,k_2)}\begin{pmatrix}
        1 & \rho_3^{(k_1,k_2)} &\dots& \rho_3^{(k_1,k_2)}\\
        \rho_3^{(k_1,k_2)} & 1 & \dots & \rho_3^{(k_1,k_2)}\\
        \vdots & & \ddots & \vdots\\
        \rho_3^{(k_1,k_2)} &\rho_3^{(k_1,k_2)} &\dots&1
    \end{pmatrix}.
    \label{cov0}
\end{equation}
Exploratory analyses of the CHEARS-AAA data indicated that measurement variability increases approximately linearly across higher frequencies \citep{curhan2018adherence},  we therefore adopt a linear parameterization to ensure model parsimony and avoid over-parameterization. Specifically, we assume $\bm{V}=\mathrm{diag}(\sigma_0 + \sigma_1, \sigma_0 + 2\sigma_1,\dots,\sigma_0 + Q\sigma_1)$, where $\sigma_0$ and $\sigma_1$ are variance parameters shared across the frequencies and groups.

\section{Estimation methods} 

The estimator for the parameters $\bm{\Theta}$ is defined as the maximum likelihood estimate of the marginal likelihood $L(\bm{\Theta};\bm{\mathcal{Y}})$ in \eqref{likelihood}. Because of the presence of latent group indicators and random effects, we employ a generalized Expectation-Maximization (EM) framework \citep{jordan1994hierarchical} based on the complete-data log-likelihood ${L}^c(\bm{\Theta};\bm{\mathcal{Y}},\bm{Z})$ in \eqref{comp-likelihood}. In contrast to standard mixture models with only within-trajectory correlations, the parameters $\bm{\varphi}_1, \dots, \bm{\varphi}_K$ cannot be estimated independently during the maximization step. The correlations between paired trajectories within a cluster induce dependence across the latent groups, requiring a joint updating procedure.

\subsection{One-stage approach}
%We first propose a one-stage estimation method that estimate parameters $\bm{\Theta}_1$ and $\bm{\Theta}_2$ simultaneously by maximizing the complete likelihood function via EM algorithm.  
 Partitioning the parameters $\bm{\Theta}$ as $\bm{\Theta}_1 = \{\bm{\varphi}, \bm{\Sigma}\}$ and $\bm{\Theta}_2 = \{\bm{\beta}, \bm{\Gamma}\}$, the conditional expectation of complete-data log-likelihood $L^c(\bm{\Theta};\bm{\mathcal{Y}},\bm{Z}|\bm{\mathcal{Y}},\hat{\bm{\Theta}}^{(\nu)})$, denoted by $U(\bm{\Theta}|\hat{\bm{\Theta}}^{(\nu)})$, can be written as $ U_1(\bm{\Theta}_1|\hat{\bm{\Theta}}^{(\nu)})  + U_2(\bm{\Theta}_2|\hat{\bm{\Theta}}^{(\nu)}) $, where
 \begin{align}
     U_1(\bm{\Theta}_1|\hat{\bm{\Theta}}^{(\nu)}) & = \sum_{i=1}^n \sum_{k_1=1}^K \sum_{k_2=1}^K \tau_{ik_1k_2}^{(\nu)}\log f_{k_1k_2}(\bm{Y}_i;(\bm{I}_2 \otimes \bm{W})\hat{\bm{\varphi}}_{k_1,k_2}^{(\nu)},\hat{\bm{\Sigma}}_{k_1k_2}^{(\nu)})   ,\label{U1} \\
     U_2(\bm{\Theta}_2|\hat{\bm{\Theta}}^{(\nu)}) & = \sum_{i=1}^n  \sum_{k_1=1}^K \sum_{k_2=1}^K  \tau_{ik_1k_2}^{(\nu)}\log \int_{\bm{b}_i}\pi_{k_1 k_2}(\bm{X}_i|\bm{b}_i;\hat{\bm{\beta}}^{(\nu)}) f(\bm{b}_i;\bm{0},\bm{\Gamma}^{(\nu)}  )\dif \bm{b}_i,
     \label{U2}
 \end{align}
 where $\tau_{ik_1k_2}^{(\nu)} = \mathrm{E}(Z_{ik_1k_2} | \bm{Y}_i,\bm{X}_i, \bm{\Theta}^{(\nu)})$.
 
The E-step evaluates the conditional expectation of  $L^c(\bm{\Theta};\bm{\mathcal{Y}},\bm{Z})$, conditional on the observed data $\bm{\mathcal{Y}}$ and the current parameter estimates $\bm{\Theta}^{(\nu)}$ at iteration $\nu$. To avoid computationally intensive integration over the random effects $\bm{b}_i$, we evaluate the mixing proportions conditional on the current empirical Bayes estimates of the random effects, $\hat{\bm{b}}_i^{(\nu)}$. Specifically, 
%In the E-step, given the observed data $\bm{\mathcal{Y}}$ and current parameter estimates $\bm{\Theta}^{(\nu)}$ at iteration $\nu$, the conditional expected latent indicator $Z_{ik_1k_2}$, denoted by $\tau_{ik_1k_2}^{(\nu)}$,   is computed as follows,
\begin{equation}
    \hat{\tau}_{ik_1k_2}^{(\nu)} =\frac{f_{k_1k_2}(\bm{Y}_i; (\bm{I}_2 \otimes \bm{W})\hat{\bm{\varphi}}_{k_1,k_2}^{(\nu)},\hat{\bm{\Sigma}}_{k_1k_2}^{(\nu)}  )\pi_{k_1k_2}(\bm{X}_i| \hat{\bm{b}}_i^{(\nu)};\hat{\bm{\beta}}^{(\nu)})}{\sum_{\ell=1}^K\sum_{h=1}^K f_{\ell h}(\bm{Y}_i;(\bm{I}_2 \otimes \bm{W})\hat{\bm{\varphi}}_{\ell,h}^{(\nu)},\hat{\bm{\Sigma}}_{\ell h}^{(\nu)})\pi_{\ell h}(\bm{X}_i|\hat{\bm{b}}_i^{(\nu)};\hat{\bm{\beta}}^{(\nu)})}, 
    \label{tau1}
\end{equation}
for $k_1,k_2=1,\dots,K; i=1,\dots,n$.  

The M-step then updates $\bm{\Theta}^{(\nu+1)}_1$ and $\bm{\Theta}^{(\nu+1)}_2$ by maximizing $U_1(\bm{\Theta}_1|\bm{\Theta}^{(\nu)})$ and $U_2(\bm{\Theta}_2|\bm{\Theta}^{(\nu)})$ separately as they are not intervened with each other in \eqref{U1} and \eqref{U2}. Specifically, by maximizing $U_1(\bm{\Theta}_1|\bm{\Theta}^{(\nu)})$ with respect to $\bm{\varphi}_k$, 
\begin{align}
    \hat{\bm{\varphi}}_k^{(\nu+1)}  &= \arg \max_{\bm{\varphi}_k} U_1(\bm{\Theta}_1|\tau_{ik_1k_2}^{(\nu)},\hat{\bm{\varphi}}^{(\nu)},\hat{\bm{\Sigma}}^{(\nu)}),\notag
\end{align} 
% This optimization reduces to the weighted least squares estimation in which a closed-form solution exists. 
for $k=1,\dots,K$. Computational details are provided  under common and group-specific covariance structures in Sections S2.1 and S2.2 of the Supplementary Material, respectively.

The maximum likelihood estimator of $\bm{\Theta}_2$ can be  obtained by maximizing $U_2(\bm{\Theta}_2|\bm{\Theta}^{(\nu)})$ defined in \eqref{U2}. However, the direct maximization of $U_2(\bm{\Theta}_2|\bm{\Theta}^{(\nu)})$is often infeasible due to the intractable integrals. To address this, numerical approximation techniques such as Gauss-Hermite quadrature, Laplace approximation and Monte Carlo methods can be applied to approximate the integrals \citep{sun2007multivariate, fu2024covariate}. While these approaches are highly effective, they can become computationally intensive when evaluated repeatedly within an iterative EM framework \citep{hedeker2006longitudinal}. As a practical alternative, we employ the unconditional maximum likelihood method of \citet{zhang2021iteratively}. This approach adapts the iteratively reweighted least squares (IRLS) algorithm \citep{green1984iteratively} for generalized linear mixed models, offering a computationally efficient option that avoids direct numerical integration while yielding competitive accuracy to the Laplace approximation. Applying this method, the updates in the M-step proceed as follows:
\begin{equation*}
\hat{\bm{\beta}}^{(\nu+1)} = \arg\max_{\bm{\beta}} U_2(\bm{\Theta}_2|\tau_{ik_1k_2}^{(\nu)},\hat{\bm{\beta}}^{(\nu)},\hat{\bm{b}}^{(\nu)}),
\end{equation*}
\begin{equation*}
\hat{\bm{b}}^{(\nu+1)}=\mathrm{E}(\bm{b}|\tau_{ik_1k_2}^{(\nu)},\hat{\bm{\beta}}^{(\nu+1)},\hat{\bm{b}}^{(\nu)}).
\end{equation*}
Computational details for estimating the exposure-phenotype coefficients $\bm{\beta}$, random effects $\bm{b}$, and its variance matrix $\bm{\Gamma}$  are provided in Section S3.1 of the Supplementary Material. 

In summary, given the initial estimates  $\hat{\bm{\Theta}}^{(0)}$ and realization $\hat{\bm{b}}^{(0)}$, the one-stage procedure iterates through the following steps until convergence. For $\nu=0,1,\dots$,

\begin{itemize}
    \item Step 1 (E-step): Compute  $\tau_{ik_1k_2}^{(\nu)}$ in \eqref{tau1} given $\hat{\bm{\Theta}}^{(\nu)}$ and $\hat{\bm{b}}^{(\nu)}$, for $k_1,k_2=1,\dots,K; i=1,\dots,n$;
    \item Step 2 (M-step): Update $\hat{\bm{\varphi}}_k^{(\nu+1)}=\arg \max_{\bm{\varphi}_k} U_1(\bm{\Theta}_1|\tau_{ik_1k_2}^{(\nu)},\hat{\bm{\varphi}}^{(\nu)},\hat{\bm{\Sigma}}^{(\nu)})$ for $k=1,\dots,K$; obtain the variance-covariance matrices $\hat{\bm{\Sigma}}^{(\nu+1)}$ based on current estimates $\tau_{ik_1k_2}^{(\nu)},\hat{\bm{\varphi}}_k^{(\nu+1)}$; update $\hat{\bm{\beta}}^{(\nu+1)} = \arg\max_{\bm{\beta}} U_2(\bm{\Theta}_2|\tau_{ik_1k_2}^{(\nu)},\hat{\bm{\beta}}^{(\nu)},\hat{\bm{b}}^{(\nu)})$ and $\hat{\bm{b}}^{(\nu+1)}=\mathrm{E}(\bm{b}|\tau_{ik_1k_2}^{(\nu)},\hat{\bm{\beta}}^{(\nu+1)},\hat{\bm{b}}^{(\nu)})$; estimate the variance matrix $\hat{\bm{\Gamma}}^{(\nu+1)}$  based on current estimates $ \hat{\bm{\beta}}^{(\nu+1)},\hat{\bm{b}}^{(\nu+1)}$.
    \end{itemize}

We found that the inner loop for updating   $\bm{\beta}$ within each M-step may require a large number of iterations to converge, consistent with observations by \cite{xu1994alternative}. To improve computational efficiency, following \cite{xu1994alternative}, we limit this inner loop at 5 iterations. Our simulation and data analysis show that this implementation does not have noticeable influence on the overall performance of the algorithm.

%To make algorithm more computationally stable,  we further propose a two-stage method that employ the multilevel mixture model with a covariate-independent mixing proportion, and in the second stage, fit a mixed-effect multinomial logit probability model using the posterior probabilities generated in the first stage. 

\subsection{Two-stage approach}
While the one-stage approach provides full maximum likelihood estimation, simultaneously estimating complex trajectory structures alongside covariate associations can lead to severe overfitting and identifiability issues when data limited relative to model complexity \citep{bolck2004estimating, fruhwirth2006finite}.  In such settings, structural covariates may inadvertently distort the formation of latent trajectory groups. 
To make estimation more stable and reliable, we propose following two-stage estimation procedure, where trajectory identification and covariate modeling are decoupled. 

In the first stage, we fit a covariate-free mixture model to reduce the model complexity.  Let $r_{k_1}$ and $r_{k_2}$ denote the averaged probabilities that two ears of all subjects are assigned to groups $k_1$ and $k_2$, respectively, for $k_1,k_2=1,\dots,K$. % Under two-stage approach, we assume that the averaged probabilities of group assignments for each ear over subjects are independent. Accordingly,  the joint mixing proportion for groups $k_1$ and $k_2$ is specified as $r_{k_1}r_{k_2}$, which is covariates-free and remains constant across subjects.
We define $\tilde{\tau}_{ik_1k_2}^{(\nu)}$ as the analogue of $\tau_{ik_1k_2}^{(\nu)}$, representing the conditional expectation of $Z_{ik_1k_2}$ conditional on current parameter estimates but with covariate-independent mixing proportions. Specifically, given the observed data $\bm{Y}_i$ and current parameters estimates $\bm{\Theta}_{1}^{(\nu)}$,  the E-step updates,  %$\tilde{\tau}_{ik_1k_2}^{(\nu)}$ is given by,
\begin{equation}
    \tilde{\tau}_{ik_1k_2}^{(\nu)} =\frac{f_{k_1k_2}(\bm{Y}_i;(\bm{I}_2 \otimes \bm{W})\hat{\bm{\varphi}}_{k_1,k_2}^{(\nu)},\hat{\bm{\Sigma}}_{k_1k_2}^{(\nu)})r_{k_1} ^{(\nu)}r_{k_2} ^{(\nu)}}{\sum_{\ell=1}^K\sum_{h=1}^K f_{\ell h}(\bm{Y}_i;(\bm{I}_2 \otimes \bm{W})\hat{\bm{\varphi}}_{\ell,h}^{(\nu)},\hat{\bm{\Sigma}}_{\ell h}^{(\nu)})r_{\ell} ^{(\nu)}r_{h} ^{(\nu)}},
    \label{tau2}
\end{equation}
for $k_1,k_2=1,\dots,K; i=1,\dots,n$. In the full joint model, directly estimating the mixing probability of a rare combinatorial pair $(k_1, k_2)$ often yields highly unstable and noisy estimates $\hat{\pi}_{k_1 k_2}$. In contrast, the marginal assignment probabilities are estimated much more stably by marginalizing over the joint assignments. By employing a factorized form for the joint group assignments in \eqref{tau2}, we stabilize the estimation of rare pairs because their joint probability mass is derived from the product of the robustly estimated marginals, $r_{k_1}$ and $r_{k_2}$. It should be noted that, although the latent group assignments are assumed independent during the E-step in the first stage, the underlying correlation between the paired trajectories is still explicitly accounted in the joint distribution in the M-step.

In the M-step, we obtain $\hat{\bm{\Theta}}_2$ by maximizing the following function:
\begin{equation}
    \tilde{U}_1(\bm{\Theta}_1|\hat{\bm{\Theta}}^{(\nu)}) = \sum_{i=1}^n \sum_{k_1=1}^K \sum_{k_2=1}^K \tilde{\tau}_{ik_1k_2}^{(\nu)}\log f_{k_1k_2}(\bm{Y}_i;(\bm{I}_2 \otimes \bm{W})\tilde{\bm{\varphi}}_{k_1,k_2}^{(\nu)},\tilde{\bm{\Sigma}}_{k_1k_2}^{(\nu)})   ,  \label{u1_twostage}
\end{equation}
for $\nu\geq1$. Specifically, 
we update
\begin{align}
r_{k_1}^{(\nu+1)}=& \sum_{i=1}^n \sum_{k_2=1}^K \tilde{\tau}_{ik_1k_2}^{(\nu)}/n, \mbox{ }
    r_{k_2}^{(\nu+1)}= \sum_{i=1}^n \sum_{k_1=1}^K \tilde{\tau}_{ik_1k_2}^{(\nu)}/n, \label{r12}
    \end{align}
    \begin{align}
\tilde{\bm{\varphi}}_k^{(\nu+1)}  = \arg \max_{\bm{\varphi}_k} \tilde{U}_1(\bm{\Theta}_1|\tilde{\tau}_{ik_1k_2}^{(\nu)},\tilde{\bm{\varphi}}^{(\nu)},\tilde{\bm{\Sigma}}^{(\nu)}).
    \label{phi2}
\end{align}
%The covariance matrices $\tilde{\bm{\Sigma}}^{(\nu)}_{k_1k_2}$ can be estimated from equations \eqref{rho} - \eqref{cov_comm2} by substituting $\tilde{\tau}_{ik_1k_2}^{(\nu)}$  for $\hat{\tau}_{ik_1k_2}^{(\nu)} $ and $\tilde{\bm{\varphi}}_k^{(\nu)}$ for $\hat{\bm{\varphi}}_k^{(\nu)}$, respectively. 
%The above procedure has a similar framework of the classical EM algorithm for the Gaussian mixture model but incorporates a modification to accommodate  the presence of distinct groups for the same participant 

In the second stage,  we  let $\tau_{ik_1k_2}^{\ast}$ denote the posterior probabilities $\tilde{\tau}_{ik_1k_2}$ obtained at the last iteration of EM algorithm in the first stage. We account for the dependence in the inference of association between assignment and exposure, and estimate the parameter vector $\bm{\Theta}_2$ 
by maximizing the following log-likelihood with posterior probabilities $\tau_{ik_1k_2}^{\ast}$:%associated with a multinomial mixed-effects logit probability model in which $\tau_{ik_1k_2}^{\ast}$ serves as the output observation:  
\begin{align}
\tilde{U}_2(\bm{\Theta}_2|\bm{\Theta}^{(t)})  = \sum_{i=1}^n  \sum_{k_1=1}^K \sum_{k_2=1}^K  \tau_{ik_1k_2}^\ast \log \int_{\bm{b}_i}\pi_{k_1 k_2}(\bm{X}_i|\bm{b}_i;\bm{\beta}^{(t)}) f(\bm{b}_i;\bm{0},\bm{\Gamma}^{(t)})\dif \bm{b}_i,
\label{u2_twostage}
\end{align} 
for $t\geq 1$. 
%In summary, given the initial estimates  $\hat{\bm{\Theta}}^{(0)}$ and the realization $\hat{\bm{b}}^{(0)}$
Estimation details for above optimization problem are provided in Section S3.2 of the Supplementary Material. In summary, the two-stage procedure consists of the following steps:
\begin{enumerate}
    \item (The first stage) Given the initial estimates  $\hat{\bm{\Theta}}_1^{(0)}$ and the realization $\hat{\bm{b}}^{(0)}$,  iterates through the following steps until convergence.  For $\nu=0,1,\dots$, 
    \begin{itemize}
        \item Step 1 (E-step): Compute  $\tilde{\tau}_{ik_1k_2}^{(\nu)}$ in \eqref{tau2} given $\tilde{\bm{\varphi}}_k^{(\nu)}$ and $\tilde{\bm{\Sigma}}^{(\nu)}$, for $k_1,k_2=1,\dots,K; i=1,\dots,n$;
        \item Step 2 (M-step): Compute $r_{k_1}^{(\nu+1)}$ and $r_{k_2}^{(\nu+1)}$  in \eqref{r12} for $k_1,k_2=1,\dots,K$; Update $\tilde{\bm{\varphi}}_k^{(\nu+1)}=\arg \max_{\bm{\varphi}_k} U_1(\bm{\Theta}_1|\tilde{\tau}_{ik_1k_2}^{(\nu)},\tilde{\bm{\varphi}}_k^{(\nu)},\tilde{\bm{\Sigma}}^{(\nu)})$ for $k=1,\dots,K$; Obtain the variance-covariance matrices $\tilde{\bm{\Sigma}}^{(\nu+1)}$ based on current estimates $\tilde{\tau}_{ik_1k_2}^{(\nu)},\tilde{\bm{\varphi}}_k^{(\nu+1)}$;
    \end{itemize}
    \item Obtain $\tau_{ik_1k_2}^{\ast}=\tilde{\tau}_{ik_1k_2}^{(\nu)}$ from the last iteration $\nu$ at the convergence of the first stage procedure;
    \item (The second stage)
Given the initial estimates  $\hat{\bm{\Theta}}_2^{(0)}$ and realization $\hat{\bm{b}}^{(0)}$,  run the following
steps until convergence:  for $t=0,1,\dots$, update $\hat{\bm{\beta}}^{(t+1)} = \arg\max_{\bm{\beta}} \tilde{U}_2(\bm{\Theta}_2|\tau_{ik_1k_2}^\ast,\hat{\bm{\beta}}^{(t)},\hat{\bm{b}}^{(t)})$ and $\hat{\bm{b}}^{(t+1)}=\mathrm{E}(\bm{b}|\hat{\bm{\beta}}^{(t+1)},\hat{\bm{b}}^{(t)})$; Estimate the variance matrix $\hat{\bm{\Gamma}}^{(t+1)}$ based on current estimates $ \hat{\bm{\beta}}^{(t+1)},\hat{\bm{b}}^{(t+1)}$.
    \end{enumerate}
%We repeat above steps until the algorithm converges and obtain the estimates of $\bm{\beta}$.

\subsection{Estimation of joint covariances}
When the correlation structure of paired trajectories is specified, the parameters in the structured variance-covariance matrix can be estimated by maximizing the expected complete log-likelihood function. For example, in CHEARS-AAA data, let $\bm{\rho}=\{\sigma_0,\sigma_1,\rho_1^{(k)},\rho_2^{(k_1,k_2)},\rho_3^{(k_1,k_2)};k,k_1,k_2=1,\dots,K\}$ denote the set of parameters of the variance-covariance matrix specified as in \eqref{cov0}. Given the current parameter estimates $\hat{\bm{\Theta}}^{(\nu)}$, the covariance parameters $\bm{\rho}$ at iteration $\nu+1$ is estimated by \begin{equation}
    \hat{\bm{\rho}}^{(\nu+1)} = \arg\max_{\bm{\rho}} U_1(\bm{\Theta}_1|\bm{\Theta}^{(\nu)}).
    \label{rho}
\end{equation}
This optimization can be performed numerically with L-BFGS-B optimization method using the standard function \texttt{optim()} in \texttt{R}. The covariance matrices $\hat{\bm{\Sigma}}_{k_1k_2}^{(\nu+1)}$  at iteration $\nu+1$ are then estimated based on current estimates $\hat{\bm{\rho}}^{(\nu+1)}$. 

Alternatively, a moment-based estimation method   \citep{xu1994alternative, jordan1995convergence} can be applied to estimate unstructured covariances. %The group-specified unstructured covariance matrices can be applied to fully capture heterogeneous data structure. 
The corresponding estimation procedure for group-specified unstructured covariance matrices is described in Section S1 of the Supplementary Material. However, when available data per group pair $(k_1,k_2)$ is limited,
the estimation of between-ear  covariance $\bm{\Sigma}_{k_1k_2}$ for all possible pairs can be unstable.
To accommodate this scenario, we propose to estimate common correlation structures shared across all group pairs to enhance model stability. Accordingly, the within-trajectory covariance matrix $\bm{\Sigma}^{(\nu+1)}_{k}$ are estimated by
\begin{align}    \hat{\bm{\Sigma}}^{(\nu+1)}_{k} = \frac{1}{2n}\sum_{i=1}^n \sum_{k=1}^K \sum_{\ell=1}^K  \hat{\tau}_{ik_1k_2}^{(\nu)} &\left[(\bm{Y}_{i1}-\bm{W}\hat{\bm{\varphi}}_k^{(\nu+1)})(\bm{Y}_{i1}-\bm{W}\hat{\bm{\varphi}}_k^{(\nu+1)})^\top \right. \notag \\
   &\left.  + (\bm{Y}_{i2}-\bm{W}\hat{\bm{\varphi}}_\ell^{(\nu+1)})(\bm{Y}_{i2}-\bm{W}\hat{\bm{\varphi}}_\ell^{(\nu+1)})^\top\right], 
    \label{cov_comm1}
\end{align}
for $k=1,\dots,K$. The between-trajectory covariance matrix $\bm{\Sigma}^{(\nu+1)}_{k_1k_2} $  are assumed to be common across all group pairs $(k_1,k_2)$ and is estimated by
\begin{align}   \hat{\bm{\Sigma}}^{(\nu+1)}_{k_1k_2}  = \frac{1}{2n}\sum_{i=1}^n \sum_{k=1}^K \sum_{\ell=1}^K \hat{\tau}_{ik_1k_2}^{(\nu)} &\left[(\bm{Y}_{i1}-\bm{W}\hat{\bm{\varphi}}_k^{(\nu+1)})(\bm{Y}_{i2}-\bm{W}\hat{\bm{\varphi}}_\ell^{(\nu+1)})^\top \right. \notag \\
    & \left. + (\bm{Y}_{i2}-\bm{W}\hat{\bm{\varphi}}_\ell^{(\nu+1)})(\bm{Y}_{i1}-\bm{W}\hat{\bm{\varphi}}_k^{(\nu+1)})^\top\right].
    \label{cov_comm2}
\end{align} 

For two-stage approach, the covariance matrices $\tilde{\bm{\Sigma}}^{(\nu)}_{k_1k_2}$ can be estimated from equations \eqref{rho} - \eqref{cov_comm2} by substituting $\tilde{\tau}_{ik_1k_2}^{(\nu)}$  for $\hat{\tau}_{ik_1k_2}^{(\nu)} $ and $\tilde{\bm{\varphi}}_k^{(\nu)}$ for $\hat{\bm{\varphi}}_k^{(\nu)}$, respectively. 
\subsection{Variance estimation of parameters of interests}
Let $\hat{\bm{\beta}}$ denote the maximum likelihood estimator of $\bm{\beta}$ for the likelihood ${L}(\bm{\Theta};\bm{\mathcal{Y}})$ in \eqref{likelihood}.  For the one-stage approach in the EM framework, the variance of $\hat{\bm{\beta}}$ can be obtained following \cite{louis1982}. Alternatively, standard errors can be estimated by directly evaluating the Hessian of the marginal observed log-likelihood ${L}(\bm{\Theta};\bm{\mathcal{Y}})$ at the final parameter estimates. Because evaluating this exact marginal likelihood requires integrating over the continuous random effects $\bm{b}_i$, we utilize a computationally efficient first-order approximation. Specifically, when the magnitudes of the linear predictor $\bm{X}_i^\top \bm{\beta}_k + b_{ik}$ and the variance of $\bm{b}_i$ are relatively small, the logistic function is approximately linear. Under these conditions, the marginal integral $\int_{\bm{b}_i} P(K_{i1}=k_1,K_{i2}=k_2|\bm{b}_i)f(\bm{b}_i)\dif \bm{b}_i$ can be reliably approximated by evaluating the conditional probability directly at the estimated conditional mean, $P(K_{i1}=k_1,K_{i2}=k_2|\hat{\mathrm{E}}[\bm{b}_i|\bm{\mathcal{Y}}])$.  %This approximation allows the observed likelihood ${L}(\bm{\Theta};\bm{\mathcal{Y}})$ and its partial derivatives with respect to $\bm{\beta}$ to be evaluated directly. %and the first- and second-order partial derivatives can be calculated empirically using the package ``pracma" in \texttt{R}. %The variance of $\hat{\bm{\beta}}$ follows from the inverse of the observed Fisher information matrix from approximated likelihood function. 
The accuracy of this approximation relies on both the small magnitudes of $\bm{X}_i \bm{\beta}_k + b_{ik}$ and variances of random effects. In scenarios where these conditions do not hold, more accurate results may be obtained using numerical integration methods, such as Gaussian quadrature, to evaluate the observed likelihood and its derivatives.

For the two-stage approach, %it follows from \cite{zhang2021iteratively} that,  when the second stage algorithm converges, the Fisher information of $\hat{\bm{\beta}}^{(t)}$ obtained at iteration $t$ of the second stage procedure converges to that of the final estimator  $\hat{\bm{\beta}}$ in probability. 
the Fisher information of $\hat{\bm{\beta}}^{(t)}$, denoted by $\bm{I}_{\bm{\beta}}^{(t)}$, is obtained by taking the negative expected values of the second-order partial derivatives of the the log-likelihood of the working response variables, following the approach of \citet{zhang2021iteratively}. This yields $\bm{I}_{\bm{\beta}}^{(t)}=(\bm{X}^\ast)^\top\bm{R}^{(t)}\bm{X}^\ast$, where $\bm{R}^{(t)}$ is the variance-covariance matrix of working responses introduced during the optimization of likelihood \eqref{u2_twostage}  (derived in Section S3 of the Supplementary Material).  The variance-covariance matrix of $\hat{\bm{\beta}}^{(t)}$ can be approximated by $\{\bm{I}_{\bm{\beta}}^{(t)}\}^{-1}$ at the convergence of the algorithm. We note that this variance estimates do not take into account the extra variation introduced by fitting  the mixture model in the first stage. To address this extra uncertainty, one could generate $B$ bootstrap samples through sampling with replacement from the original dataset. Following \citet{ren2010nonparametric}, only the highest hierarchical level (i.e., participants, in the context of our AAA example) should be re-sampled when bootstrapping clustered data. For each bootstrap sample $b=1,\dots,B$, the entire two-stage procedure is repeated to estimate $\hat{\bm{\beta}}_b$ from the likelihood in \eqref{u2_twostage}. The bootstrap variance estimate is then computed as the sample variance of $\{\hat{\bm{\beta}}_b:b=1,\dots,B\}$.

\subsection{Determination of the number of groups}
Standard penalized likelihood criteria, such as the Bayesian Information Criterion (BIC) and Akaike Information Criterion (AIC), are commonly used to select the number of trajectory groups in GBTM \citep{nagin2018group}. %The properties of various fit statistics for GBTM have been further investigated in  \citep{klijn2017introducing}. % It is also noted by \citet{nagin2018group} that, instead of relying on the fit statistic, the selection of the number of trajectory groups should be evaluated on model's substantive interest. For instance, 
 In this paper, we apply BIC to select the number of groups for our proposed mixture model. Specifically,
 \begin{equation*}
     \mbox{BIC}=\log(L(\bm{\Theta};\bm{\mathcal{Y}}))-0.5m\log(n),
 \end{equation*}
 where ${L}(\bm{\Theta};\bm{\mathcal{Y}})$ is the observed likelihood as specified in \eqref{likelihood} obtained from the model, $m$ is the total number of parameters estimated in the model.

 \section{Simulation studies}
We evaluated the proposed model through simulation studies that mimic the CHEARS-AAA data. We assumed the number of phenotypes $K=4$, and 
set the following true parameters as $(\beta_{02},\beta_{03}, \beta_{04}, \beta_{12},\beta_{13}, \beta_{14})=(-\log(0.5), 0.5, \log(1.5), 1.5, \log(2), 1)$. We generated $\pi_{k}(\bm{X}_{ij}|\bm{b}_i)$, the probability of assigning to phenotype $k$ for ear $j$ of participant $i$, using mixed-effect multinomial logistic regression model
\begin{equation*}
    \log \left(\frac{\pi_{k}(\bm{X}_{ij}|\bm{b}_i)}{\pi_{1}(\bm{X}_{ij}|\bm{b}_i)}\right) = \bm{X}_{ij}^\top\bm{\beta}_k+b_{ik}=\beta_{0k}+X_{ij}\beta_{1k}+b_{ik}, \mbox{ } k= 2,3,4.
\end{equation*}
%where $k=1$ was taken as the reference and the probabilities of belonging to $K$ phenotypes sum to 1. 
Specifically,  $P_{ijk}=\exp (\bm{x}_{ij}^\top\bm{\beta}_k+b_{ik})/\left[1+\sum_{l=2}^{K} \exp (\bm{x}_{ij}^\top\bm{\beta}_l+b_{il})\right]$ for $k=2,3,4$, and $P_{ij1}=1/\left[1+\sum_{l=2}^{K} \exp (\bm{x}_{ij}^\top\bm{\beta}_l+b_{il})\right]$.  The covariates $X_{i1}=X_{i2}$ were generated independently from $N(0,0.5)$. The random effects $b_{ik}$ were generated independently from  $N(0, \gamma_k)$ with true parameter $(\gamma_1,\gamma_2,\gamma_3,\gamma_4)=(0.1,0.2,0.2,0.3)$. 
Consequently, the phenotypes $K_{ij}$ for $j$th ear of $i$th subject were generated from a categorical distribution with 4 phenotypes and each with probability $P_{ijk}$ for $k=1,\dots,4$. Let $\bm{\varphi}_k=(\varphi_{0k},\varphi_{1k},\varphi_{2k})^\top$ denote the parameters for phenotype $k$ associated with frequency thresholds in mixture model, for $k=1,\dots,4$. We set the following true parameters as $\bm{\varphi}_1=(1.8,1.8,0.5)^\top, \bm{\varphi}_2=(11.7,2.1,0.8)^\top,\bm{\varphi}_3=(4.6,5.5,3.2)^\top,\bm{\varphi}_4=(18.5,3.9,4.8)^\top$.
 The error terms $\bm{\varepsilon}_i=(\varepsilon_{i11},\dots,\varepsilon_{i1Q},\varepsilon_{i21},\dots,\varepsilon_{i2Q})^\top$ were generated from multivariate normal with zero means and covariance matrix \eqref{sigma}. %The standard deviation $\bm{V}=\mathrm{diag}(\sigma_0 + \sigma_1, \sigma_0 + 2\sigma_1,\dots,\sigma_0 + Q\sigma_1)$, and
 We adopted following exchangeable correlation structure,
\begin{equation}     \bm{\Lambda}_{k_1}=\bm{\Lambda}_{k_2}=\begin{pmatrix}
        1 & \rho_1 &\dots& \rho_1\\
        \rho_1 & 1 & \dots & \rho_1\\
        \vdots & & \ddots & \vdots\\
        \rho_1 &\rho_1 &\dots&1
    \end{pmatrix}, \mbox{ } \bm{C}_{k_1k_2}=\begin{pmatrix}
        \rho_2 & \rho_2\rho_3 &\dots& \rho_2\rho_3\\
        \rho_2\rho_3 & \rho_2 & \dots & \rho_2\rho_3\\
        \vdots & & \ddots & \vdots\\
        \rho_2\rho_3 &\rho_2\rho_3 &\dots&\rho_2
    \end{pmatrix}.
    \label{ex}
\end{equation} 
For design matrix $\bm{W}$, we employed a quadratic form  based on the frequency thresholds $h_1, \dots,  h_Q$ corresponding to $Q=7$ frequencies assessed. Specifically,
\begin{equation}
    \bm{W}= \begin{pmatrix}
        1 & h_1 & h^2_1 \\
        1 & h_2 & h_2^2 \\
        \vdots & \vdots &\vdots \\
        1 & h_Q & h_Q^2
    \end{pmatrix},
    \label{designmat}
\end{equation}
where the frequency thresholds were set to $(h_1,\dots,h_Q) = (500, 1000, 2000, 3000, 4000, 6000, 8000)$ Hz and scaled by dividing by 2000, resulting in $(h_1,\dots,h_Q) = (0.25, 0.5, 1, 1.5, 2, 3, 4)$.
Conditional on $K_{ij}$, the paired hearing threshold data $\bm{Y}_i$ were generated using the marginal model $Y_{ijq}|_{K_{ij}=k}=\varphi_{0k}+\varphi_{1k}h_q+\varphi_{2k}h_q^2+\varepsilon_{ijq}$. 

We explored two scenarios reflecting the variability of standard deviations across frequencies for both ears. The first setting features the greater variation observed  in the CHEARS-AAA data, where we set $\sigma_0 = 4$ and $\sigma_1 = 3$, yielding standard deviations $\bm{V}=\mathrm{diag}(7, 10, 13, 16, 19, 22, 25)$  across seven frequencies.  The second setting examines lower variability, with $\sigma_0 = 0.8$ and $\sigma_1 = 1.2$, resulting in   $\bm{V}=\mathrm{diag}(2.0, 3.2, 4.4, 5.6, 6.8, 8.0, 9.2)$. We fixed the within-ear correlation at $\rho_1 = 0.5$ and the between-ear different-frequency reduction parameter at $\rho_3 = 0.6$. To evaluate model performance under varying levels of between-ear correlation, we considered three values for $\rho_2$: 0 (no correlation), 0.5 (moderate), and 0.8 (high). For each scenario, we compared the proposed one-stage (OneStg) and two-stage (TwoStg) estimation procedures to conventional GBTM (IndGBTM). IndGBTM was implemented under one-stage framework that commonly adopted in conventional GBTMs, but does not account for the correlation between trajectories from the two ears of the same participant. We considered proposed model with both exchangeable and unstructured correlation matrices using a common covariance matrix shared across all groups. %As the simulation results under exchangeable and unstructured correlation  were similar, we report findings for the exchangeable correlation model in the main text, with results for unstructured correlation matrices provided in Tables 1 and 2 in the Supplementary Materials.
To evaluate simulation performance, we reported relative bias (RB), empirical standard error (ESE), analytical standard error (ASE), and empirical coverage probability (CP) of the slope parameters $\beta_{12},\beta_{13},\beta_{14}$, each averaged over 500 replications. RB was computed as the mean of $(\hat{\beta}_{1j} - \beta_{1j}) / \beta_{1j}$ for $j = 1, 2, 3$. ESE was  defined as the standard deviation across replications, and ASE was obtained from the observed likelihood for one-stage methods and by bootstrap approach for the two-stage method with
the bootstrap sample sizes of 200.   % Since our primary focus is on the slope parameters, results for the intercept parameters $\beta_{02}, \beta_{03}, \beta_{04}$ are provided in Tables 3 and 4 of the Supplementary Materials.

\begin{table*}
\caption{Relative bias (RB), averaged standard error (ASE), empirical standard error (ESE), and coverage probability (CP) of the parameter estimates of interest following the proposed model based on 500 replications. The covariance matrix of paired trajectories was estimated from maximum likelihood under the exchangeable structure, with $\sigma_0=4,\sigma_1=3$.}
\label{table1}
\centering
\begin{tabular}{@{}lclrrrrrrrrrrrr@{}}
\hline
 &  &  & \multicolumn{4}{c}{$\beta_{12}$} & \multicolumn{4}{c}{$\beta_{13}$} & \multicolumn{4}{c}{$\beta_{14}$} \\
\cline{4-7}\cline{8-11}\cline{12-15}
$N$ & $\rho_2$ & Methods & RB & ESE & ASE & CP & RB & ESE & ASE & CP & RB & ESE & ASE & CP \\
\hline
200 & 0   & IndGBTM & $-$0.078 & 0.316 & 0.356 & 95.2 & $-$0.104 & 0.321 & 0.349 & 95.6 & $-$0.082 & 0.333 & 0.357 & 94.7 \\
    &     & OneStg  & $-$0.080 & 0.315 & 0.355 & 95.1 & $-$0.103 & 0.320 & 0.349 & 95.6 & $-$0.082 & 0.334 & 0.357 & 94.5 \\
    &     & TwoStg  & $-$0.072 & 0.310 & 0.333 & 93.5 & $-$0.057 & 0.307 & 0.337 & 96.5 & $-$0.060 & 0.330 & 0.347 & 94.8 \\
[0.3em] 
    & 0.5 & IndGBTM & $-$0.085 & 0.318 & 0.359 & 94.2 & $-$0.089 & 0.316 & 0.354 & 95.4 & $-$0.084 & 0.343 & 0.361 & 93.5 \\
    &     & OneStg  & $-$0.059 & 0.314 & 0.356 & 94.6 & $-$0.049 & 0.313 & 0.353 & 96.2 & $-$0.054 & 0.340 & 0.360 & 94.2 \\
    &     & TwoStg  & $-$0.049 & 0.311 & 0.336 & 95.2 & $-$0.025 & 0.311 & 0.339 & 96.0 & $-$0.037 & 0.332 & 0.348 & 94.6 \\
[0.3em] 
    & 0.8 & IndGBTM & $-$0.090 & 0.323 & 0.359 & 92.9 & $-$0.098 & 0.320 & 0.354 & 95.2 & $-$0.089 & 0.346 & 0.360 & 94.2 \\
    &     & OneStg  & $-$0.059 & 0.316 & 0.355 & 95.0 & $-$0.049 & 0.316 & 0.353 & 96.5 & $-$0.054 & 0.341 & 0.360 & 94.8 \\
    &     & TwoStg  & $-$0.044 & 0.318 & 0.336 & 94.8 & $-$0.017 & 0.312 & 0.340 & 96.2 & $-$0.032 & 0.338 & 0.348 & 94.8 \\
[0.3em] 
500 & 0   & IndGBTM & $-$0.088 & 0.229 & 0.218 & 88.8 & $-$0.100 & 0.218 & 0.214 & 91.7 & $-$0.082 & 0.220 & 0.219 & 92.1 \\
    &     & OneStg  & $-$0.088 & 0.229 & 0.217 & 89.0 & $-$0.105 & 0.218 & 0.213 & 91.0 & $-$0.086 & 0.220 & 0.218 & 92.1 \\
    &     & TwoStg  & $-$0.077 & 0.217 & 0.206 & 89.8 & $-$0.070 & 0.212 & 0.209 & 93.3 & $-$0.061 & 0.210 & 0.213 & 94.2 \\
[0.3em] 
    & 0.5 & IndGBTM & $-$0.092 & 0.228 & 0.218 & 87.5 & $-$0.107 & 0.218 & 0.214 & 91.9 & $-$0.087 & 0.222 & 0.218 & 91.2 \\
    &     & OneStg  & $-$0.064 & 0.225 & 0.215 & 91.5 & $-$0.063 & 0.216 & 0.213 & 93.3 & $-$0.056 & 0.219 & 0.218 & 94.2 \\
    &     & TwoStg  & $-$0.052 & 0.217 & 0.209 & 92.9 & $-$0.033 & 0.213 & 0.210 & 93.5 & $-$0.035 & 0.212 & 0.215 & 94.6 \\
[0.3em] 
    & 0.8 & IndGBTM & $-$0.091 & 0.233 & 0.218 & 87.3 & $-$0.105 & 0.222 & 0.214 & 91.2 & $-$0.086 & 0.224 & 0.219 & 90.6 \\
    &     & OneStg  & $-$0.061 & 0.225 & 0.216 & 91.7 & $-$0.057 & 0.218 & 0.214 & 91.7 & $-$0.052 & 0.219 & 0.218 & 93.3 \\
    &     & TwoStg  & $-$0.047 & 0.218 & 0.208 & 92.9 & $-$0.027 & 0.213 & 0.210 & 93.1 & $-$0.031 & 0.213 & 0.215 & 94.8 \\
\hline
\end{tabular}
\end{table*}

\begin{table*}
\caption{Relative bias (RB), averaged standard error (ASE), empirical standard error (ESE), and coverage probability (CP) of the parameter estimates of interest following the proposed model based on 500 replications. The covariance matrix of paired trajectories was estimated using the moment-based estimation under the unstructured correlation structure, with $\sigma_0=4,\sigma_1=3$.}
\label{table2}
\centering
\begin{tabular}{@{}lclrrrrrrrrrrrr@{}}
\hline
 &  &  & \multicolumn{4}{c}{$\beta_{12}$} & \multicolumn{4}{c}{$\beta_{13}$} & \multicolumn{4}{c}{$\beta_{14}$} \\
\cline{4-7}\cline{8-11}\cline{12-15}
$N$ & $\rho_2$ & Methods & RB & ESE & ASE & CP & RB & ESE & ASE & CP & RB & ESE & ASE & CP \\
\hline
200 & 0   & IndGBTM & $-$0.090 & 0.317 & 0.357 & 94.2 & $-$0.109 & 0.323 & 0.352 & 95.2 & $-$0.094 & 0.333 & 0.359 & 95.0 \\
    &     & OneStg  & $-$0.091 & 0.320 & 0.356 & 94.0 & $-$0.112 & 0.321 & 0.351 & 95.2 & $-$0.104 & 0.372 & 0.358 & 94.4 \\
    &     & TwoStg  & $-$0.071 & 0.314 & 0.340 & 94.0 & $-$0.058 & 0.310 & 0.340 & 95.8 & $-$0.061 & 0.328 & 0.350 & 94.4 \\[0.3em]
    & 0.5 & IndGBTM & $-$0.084 & 0.318 & 0.358 & 93.8 & $-$0.087 & 0.315 & 0.353 & 95.8 & $-$0.082 & 0.342 & 0.360 & 93.3 \\
    &     & OneStg  & $-$0.059 & 0.314 & 0.355 & 94.4 & $-$0.049 & 0.314 & 0.353 & 96.5 & $-$0.054 & 0.341 & 0.360 & 94.0 \\
    &     & TwoStg  & $-$0.049 & 0.312 & 0.340 & 95.4 & $-$0.025 & 0.310 & 0.341 & 96.2 & $-$0.037 & 0.334 & 0.350 & 94.0 \\[0.3em]
    & 0.8 & IndGBTM & $-$0.088 & 0.322 & 0.349 & 92.1 & $-$0.093 & 0.314 & 0.349 & 96.0 & $-$0.086 & 0.336 & 0.356 & 94.2 \\
    &     & OneStg  & $-$0.058 & 0.317 & 0.354 & 95.0 & $-$0.047 & 0.317 & 0.353 & 96.5 & $-$0.053 & 0.342 & 0.360 & 94.6 \\
    &     & TwoStg  & $-$0.043 & 0.319 & 0.339 & 94.6 & $-$0.016 & 0.312 & 0.341 & 96.0 & $-$0.031 & 0.339 & 0.349 & 94.2 \\[0.3em]
500 & 0   & IndGBTM & $-$0.086 & 0.228 & 0.218 & 89.0 & $-$0.097 & 0.218 & 0.214 & 91.7 & $-$0.081 & 0.219 & 0.219 & 92.7 \\
    &     & OneStg  & $-$0.086 & 0.228 & 0.217 & 89.2 & $-$0.103 & 0.218 & 0.213 & 91.5 & $-$0.084 & 0.219 & 0.218 & 92.5 \\
    &     & TwoStg  & $-$0.076 & 0.217 & 0.207 & 89.8 & $-$0.069 & 0.212 & 0.209 & 92.7 & $-$0.060 & 0.210 & 0.214 & 94.2 \\[0.3em]
    & 0.5 & IndGBTM & $-$0.090 & 0.228 & 0.217 & 87.5 & $-$0.103 & 0.218 & 0.214 & 92.1 & $-$0.085 & 0.222 & 0.218 & 91.9 \\
    &     & OneStg  & $-$0.064 & 0.225 & 0.215 & 90.6 & $-$0.061 & 0.216 & 0.213 & 92.9 & $-$0.055 & 0.219 & 0.218 & 93.3 \\
    &     & TwoStg  & $-$0.051 & 0.218 & 0.210 & 92.7 & $-$0.032 & 0.213 & 0.211 & 93.3 & $-$0.035 & 0.212 & 0.215 & 94.8 \\[0.3em]
    & 0.8 & IndGBTM & $-$0.089 & 0.221 & 0.212 & 88.5 & $-$0.101 & 0.212 & 0.211 & 92.9 & $-$0.084 & 0.215 & 0.216 & 92.5 \\
    &     & OneStg  & $-$0.060 & 0.225 & 0.216 & 91.2 & $-$0.055 & 0.218 & 0.213 & 91.9 & $-$0.051 & 0.219 & 0.218 & 94.0 \\
    &     & TwoStg  & $-$0.047 & 0.218 & 0.209 & 92.7 & $-$0.026 & 0.213 & 0.210 & 92.9 & $-$0.030 & 0.213 & 0.215 & 95.0 \\
\hline
\end{tabular}
\end{table*}
Table~\ref{table1} summarizes the estimation performance of the IndGBTM, OneStg, and TwoStg methods for   $\beta_{12},\beta_{13},\beta_{14}$ in the presence of high within-trajectory measurement variability, assuming an exchangeable correlation structure described in \eqref{ex}. 
    We observed that, when $\rho_2=0$, all three methods demonstrated similar levels of bias and coverage near the nominal level of $95\%$, indicating the proposed methods performed well even when trajectories were independent. However, as $\rho_2$ increases above zero, both the OneStg and TwoStg methods outperformed IndGBTM, demonstrating lower RBs and improved CPs that were closer to the nominal level. These improvements became more evidence as between-ear correlation strengthened, particularly for the TwoStg approach, which consistently yielded the lowest RBs and CPs closest to the nominal especially with larger sample sizes. For all methods, ESE and ASE were close to each other, and
CPs were generally close to the nominal $95\%$ level across most scenarios, while IndGBTM tended to exhibit slightly lower CPs, especially as $\rho_2$ increased. The TwoStg approach typically exhibited the smallest ESE among the three methods.  Table~\ref{table2} presents the estimation results under an unstructured covariance and shows similar patterns to those observed in Table~\ref{table1}. Additionally,
when within-trajectory variance was low, all three methods exhibited comparable and minimal RB and CP near the nominal $95\%$ level for $\hat{\beta}_{21}$, $\hat{\beta}_{31}$, $\hat{\beta}_{41}$ (see Tables~S1--S2 in the Supplementary Material). As the effect of ignoring between-ear correlation becomes negligible when measurement variability is small, it is not unexpected that the proposed methods did not demonstrate a significant improvement over IndGBTM. To  summarize, our simulation results suggest that by accounting for hierarchical correlation, the proposed OneStg and TwoStg methods provide more accurate estimations than IndGBTM when between-ear correlation is present, and perform comparably to IndGBTM when such correlation is negligible or measurement error variance is minimal.

\section{Analysis of the CHEARS-AAA data}
In this section, we applied our method to the CHEARS-AAA data.  We examined the association between the Dietary Approaches to Stop Hypertension (DASH) dietary adherence score and audiometric phenotypes of age-related hearing loss. A higher DASH score indicated greater adherence to the recommendations provided for the DASH diet. Previous studies \citep{curhan2018adherence, curhan2020prospective} demonstrated that  greater adherence to the DASH dietary pattern \citep{appel2006dietary} was associated with a substantially lower risk of hearing loss in this cohort.   Following  \citet{curhan2018adherence,curhan2020prospective}, we categorized DASH scores into quintiles (Q1-Q5). A review of participant characteristics across DASH quintiles is given in Table S3 of the Supplementary Material. Participants with missing values for specific hearing thresholds or covariates (approximately $2\%$) were excluded. The data for analysis included 3047 participants, each of whom had hearing assessments for both ears.

In the analysis of the CHEARS-AAA data, mean trajectories were modeled using the design matrix $\mathbf{W}$ specified in \eqref{designmat}, and the probabilities of phenotype assignment were modeled using a mixed-effects multinomial logistic regression, i.e.,
\begin{equation}
    \log \left(\frac{\pi_{k}(\bm{X}_{ij}|\bm{b}_i)}{\pi_{1}(\bm{X}_{ij}|\bm{b}_i)}\right) =\beta_{0k}+\beta_{1k}I(\mbox{DASH}_i=2)+\dots+\beta_{4k}I(\mbox{DASH}_i=5) + \dots+ b_{ik},
    \label{logitreal}
\end{equation}
for ear $j=1,2$ of participant $i=1,\dots,3047$ and phenotype $k=1,\dots,K$, where the lowest DASH quintile served as the reference and the probabilities of belonging to $K$ phenotypes sum to 1. Model \eqref{logitreal} also includes additional confounders that are not shown explicitly. These included age (continuous), self-reported major ancestry (binary: white, other), smoking history (binary: never, ever), body mass index (BMI) (continuous), total energy intake (continuous in calories), persistent tinnitus (several days per week or daily; binary: yes or no), and physical activity level (categorical: 1--5; representing quintiles). Sex was not included as a confounder as all study participants were female. The exchangeable structure specified in \eqref{ex} and the unstructured covariance structure were considered. \citet{dubno2013classifying} classified audiograms of older adults into four    phenotypes (older-normal, metabolic, sensory, and metabolic plus sensory). Guided by this substantive classification, we evaluate the proposed models with two to four patterns in our analysis  to balance interpretability with reference to prior research.

 When applying one-stage approach to the CHEARS-AAA dataset, the predicted phenotype 1 was assigned to over 70\% of ears, while phenotypes 2 and 4 accounted for only 6.8\% and 5.3\%, respectively (see Table S4 in Supplementary Material). This notable group size imbalance results in a limited effective sample size for one-stage approach to successfully distinguish the rare groups. Given this reason, the one-stage approach experienced identifiability challenge when applied to the CHEARS-AAA data. Hence, we focused our primary analysis on the two-stage approach, which offered more stable and interpretable results in this context.  Results are presented based on the exchangeable covariance structure with four phenotypes in Table~\ref{dash}, which provided the best fit according to BIC when compared to models with an unstructured covariance structure or different numbers of phenotypes (two or three). %Alternative choices of covariance structure resulted in different group assignments and fitted shapes. 
See Figures S1 and Tables~S5--S6 in the Supplementary Material for results under an unstructured covariance structure. To account for the variation introduced in the first stage of two-stage approach, we estimate the standard deviations of parameter estimates using bootstrapping. Following \cite{yang2024soft}, only the highest level units (participants) in the CHEARS-AAA data were sampled with replacement. Lower level data, including ears and hearing measurements of sampled participants, were not resampled, as additional resampling at these levels does not improve bootstrap performance \citep{ren2010nonparametric}. The proposed model was evaluated using 200 bootstrap samples of the CHEARS-AAA dataset, and  bootstrap confidence intervals, as well as those following from the Fisher information in the second stage procedure, were reported in Table~\ref{dash}.

\begin{table*}
\caption{Odds ratios and $95\%$ confidence intervals for the association of DASH quintiles (Q2--Q5) with specific phenotypes relative to the lowest quintile (reference Q1), estimated from the proposed two-stage approach under the exchangeable structure based on the CHEARS-AAA dataset. Confidence intervals derived from Fisher information are reported in round brackets and confidence intervals from bootstrap standard error estimates are reported in square brackets.}
\label{dash}
\centering
\begin{tabular}{@{}lccc@{}}
\hline
\texttt{DASH} & Phenotype 2 & Phenotype 3 & Phenotype 4 \\
\hline
Q2 & 1.03 (0.99, 1.07) [0.66, 1.40] & 1.08 (1.06, 1.10) [0.83, 1.33] & 0.92 (0.86, 0.98) [0.10, 1.74] \\
Q3 & 0.84 (0.80, 0.88) [0.43, 1.25] & 0.90 (0.88, 0.92) [0.63, 1.17] & 0.78 (0.72, 0.84) [0.17, 1.39] \\
Q4 & 1.06 (1.02, 1.10) [0.65, 1.47] & 0.88 (0.86, 0.90) [0.59, 1.17] & 0.87 (0.81, 0.93) [0.24, 1.50] \\
Q5 & 1.05 (0.99, 1.11) [0.60, 1.50] & 0.77 (0.73, 0.81) [0.48, 1.06] & 0.89 (0.83, 0.95) [0.05, 1.73] \\
\hline
\end{tabular}
\end{table*}

\begin{figure}
    \centering
    \includegraphics[width=0.7\linewidth]{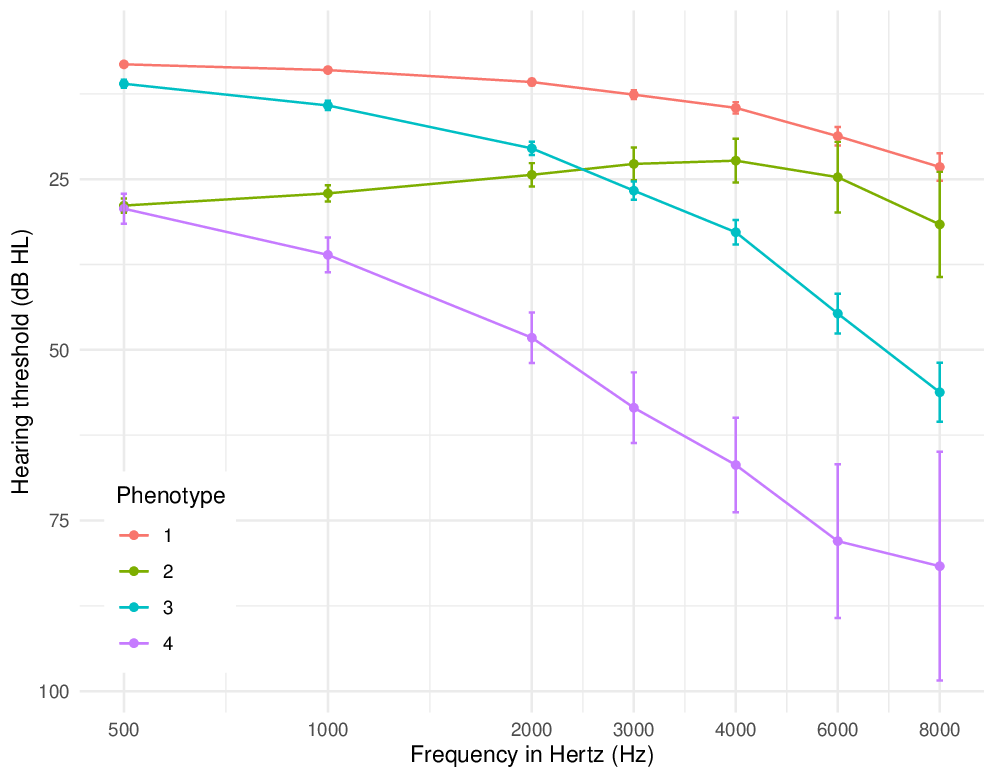}
    \caption{Predicted average audiometric patterns (hearing thresholds in dB across frequencies in Hz) and 95\% confidence intervals for phenotypes identified by the proposed GBTM with exchangeable covariance structure on the CHEARS-AAA dataset. The number and percentage of ears in each phenotype are as follows: phenotype 1, 4486 (73.6\%); phenotype 2, 415 (6.8\%); phenotype 3, 869 (14.3\%); and phenotype 4, 324 (5.3\%).}
    \label{real_traj}
\end{figure}

 Table \ref{dash} presents odds ratios and corresponding confidence intervals for hearing phenotypes 2, 3, and 4, reflecting the odds of having the specific phenotype compared to the reference category phenotype 1. In this analysis,  DASH quintile Q1 indicates a diet that  least reflects adherence to the DASH dietary pattern, and Q5 indicates a diet that most resembles the DASH dietary pattern. The participant characteristics according to phenotypes 1--4 are presented in Table S4 in the Supplementary Material. The results in Table \ref{dash} suggested that, participants with greater adherence to the DASH dietary pattern (i.e., those in higher DASH quintiles) tended to have lower odds of exhibiting phenotype 3 or 4 compared to those in the lowest quintile, with the lowest odds observed in the highest quintile.  Additionally, Figure~\ref{real_traj} shows predicted
averaged audiometric patterns by the groups that were determined using the proposed method. The audiogram displays hearing thresholds, (i.e. the lowest sound intensity or level that a person can detect at a specific frequency), on a graph.  The Y-axis shows intensity (dB) and the X-axis shows the sound frequency (Hz). Lower thresholds (lower decibel values at the top) indicate better hearing, while higher thresholds (dB values) at the bottom indicate worse hearing. From Figure~\ref{real_traj}, Phenotype 3 differs from phenotype 2 with respect to hearing threshold values in the lower frequencies and the slope of the audiogram between 2000 and 6000 Hz. Specifically, phenotype 3 exhibits lower thresholds and a steeper slope than phenotype 2. The lower frequency thresholds were similar for phenotypes 3 and 1. Phenotypes 3 and 4 showed similar higher frequency slopes, while phenotype 4 is characterized by overall poorer thresholds across frequencies. These groupings are similar with those identified in prior work \citep{dubno2013classifying} that characterized on four audiometric phenotypes.

\section{Discussion}
Traditional GBTM relies on the assumption of conditional independence, positing that all structural patterns are explained by the latent class, leaving only independent residual errors. However, in the context of repeated measurements from paired organs (e.g., left and right ears), this assumption can be biologically unrealistic. Our proposed model relaxes this conditional independence assumption and captures the inherent hierarchical dependence between paired trajectories. It is important to note that, because we consider the covariance structure to the residual errors rather than introducing individual-level random effects into the mean trajectory, we successfully account for multilevel correlation while preserving the strict, taxonomic philosophy of GBTM. 

The proposed model is flexible and can accommodate various specifications of the design matrix of mean trajectories, including cubic polynomials or spline-based methods to further enhance model flexibility. Simulations demonstrated that ignoring hierarchical dependence can lead to biased estimation and  inference. When the correlation between trajectories within a cluster is negligible, the conventional GBTM serves an appropriate and robust analytic choice. Nevertheless, even when the correlation across trajectories within clusters is not apparent, our proposed methods demonstrate performance comparable to standard GBTM. %While existing standard software (e.g., \texttt{PROC TRAJ} and the \texttt{lcmm} package) can fit multivariate trajectory models, these tools generally enforce conditional independence between outcomes or force paired trajectories into a single shared latent class.
To facilitate the implementation of these models, we have developed the \texttt{R} package \texttt{pairtraj}, available at \url{https://github.com/yxc170/pairtraj}.

We developed two likelihood-based estimating procedures: a statistically efficient one-stage approach and a computationally robust two-stage approach. % we developed the \texttt{R} package \texttt{pairgbtm} to facilitate the implementation of the proposed model.
 %While adding a covariance matrix is a recognized concept in general mixture modeling, its specific formulation for paired trajectories is methodologically challenging and absent from standard trajectory software. 
%: (1) the one-stage EM algorithm that jointly estimates population trajectory pattern and exposure effects on trajectory group memberships, and (2) the two-stage procedure that offers a more numerically stable alternative by first fitting a covariate-free mixture to obtain group membership assignments probabilities, then linking these probabilities to exposures via a mixed-effects multinomial logistic model.  
%For the corresponding \texttt{R} code, please visit https://github.com/yxc170/pairgbtm.
The proposed one-stage approach offers a preferable statistical efficiency by jointly estimating mixture components and mixing probabilities. However, 
when group sizes were small or highly imbalanced,  two-stage approach is highly recommended.  
Although both estimation approaches can be affected by small group sizes,  the two-stage approach  reduces parameter complexity and produces more stable group assignments by fitting
a covariate-free mixture in the first stage, and facilitates more reliable estimation of exposure effects on group membership in the second stage. It is also worth noting that there is a trade-off between the flexibility of correlation structure and the computational stability of the mixture model. For instance, in the CHEARS-AAA analysis, group-specific unstructured covariance matrices can better capture heterogeneity in the data, but the large number of covariance parameters can make the regression parameters poorly determined by the data, particularly for the rare group. 

Although our method was illustrated in the context of the CHEARS-AAA dataset with two levels of clustering by ears and participants, the proposed framework can be extended to accommodate more complex hierarchical designs, such as further longitudinal clustering. Future work will focus on extending the model to longitudinal data involving multiple time points by incorporating additional random effects for time $t$ in the nonlinear mixed effects model for the group
membership, and by allowing the standard deviation of mean trajectories to vary by time.  Moreover, the correlation structure can be generalized to higher dimensions to capture additional dependence introduced by repeated measurements over time. By augmenting the relevant vectors and matrices, the likelihood function and maximum likelihood estimation procedures can be modified to accommodate these extensions.

%%%%%%%%%%%%%%%%%%%%%%%%%%%%%%%%%%%%%%%%%%%%%%
%% Support information, if any,             %%
%% should be provided in the                %%
%% Acknowledgements section.                %%
%%%%%%%%%%%%%%%%%%%%%%%%%%%%%%%%%%%%%%%%%%%%%%
\begin{acks}[Acknowledgments]
{\it Conflict of Interest}: None declared. 
\end{acks}

%%%%%%%%%%%%%%%%%%%%%%%%%%%%%%%%%%%%%%%%%%%%%%
%% Funding information, if any,             %%
%% should be provided in the                %%
%% funding section.                         %%
%%%%%%%%%%%%%%%%%%%%%%%%%%%%%%%%%%%%%%%%%%%%%%
\begin{funding}
This work was supported by National Institute Health grants [R01 DC017717 and U01 CA176726 (NHS II)].
\end{funding}

%%%%%%%%%%%%%%%%%%%%%%%%%%%%%%%%%%%%%%%%%%%%%%
%% Supplementary Material, including data   %%
%% sets and code, should be provided in     %%
%% {supplement} environment with title      %%
%% and short description. It cannot be      %%
%% available exclusively as external link.  %%
%% All Supplementary Material must be       %%
%% available to the reader on Project       %%
%% Euclid with the published article.       %%
%%%%%%%%%%%%%%%%%%%%%%%%%%%%%%%%%%%%%%%%%%%%%%
\begin{supplement}
\stitle{Supplementary Material for ``Joint Group-Based Trajectory Modeling for Paired Repeated Measures: An Application to Audiometric Phenotypes and Dietary Associations"}
\sdescription{The Supplementary material provides technical derivations of the results and additional empirical results unshown in the manuscript, including extra simulations, detailed CHEARS-AAA participant characteristics, phenotype-specific audiometric patterns, and DASH--phenotype association estimates under unstructured correlation matrix.}
\end{supplement}

%%%%%%%%%%%%%%%%%%%%%%%%%%%%%%%%%%%%%%%%%%%%%%%%%%%%%%%%%%%%%
%%                  The Bibliography                       %%
%%                                                         %%
%%  imsart-nameyear.bst  will be used to                   %%
%%  create a .BBL file for submission.                     %%
%%                                                         %%
%%  Note that the displayed Bibliography will not          %%
%%  necessarily be rendered by Latex exactly as specified  %%
%%  in the online Instructions for Authors.                %%
%%                                                         %%
%%  MR numbers will be added by VTeX.                      %%
%%                                                         %%
%%  Use \cite{...} to cite references in text.             %%
%%                                                         %%
%%%%%%%%%%%%%%%%%%%%%%%%%%%%%%%%%%%%%%%%%%%%%%%%%%%%%%%%%%%%%

%% if your bibliography is in bibtex format, uncomment commands:
\bibliographystyle{imsart-nameyear} % Style BST file
%\bibliography{bibliography}       % Bibliography file 
\bibliography{refs} 
%(usually '*.bib')

%% or include bibliography directly:

\end{document}